# Additive and Subtractive Color Filters Based on Birefringent Dielectric Metasurfaces

Ondřej Červinka[1], Filip Ligmajer[1,2], Ondřej Brunn[3], Miroslav Horáček[3], Stanislav Krátký[3], Tomáš Šikola[1,2,*]

[1]Brno University of Technology, Central European Institute of Technology, Purkyňova 123, 61200 Brno, Czech Republic

[2]Brno University of Technology, Faculty of Mechanical Engineering, Institute of Physical Engineering, Technická 2, 61669, Brno, Czech Republic

[3]Institute of Scientific Instruments, Czech Academy of Sciences, Královopolská 147, 612 64 Brno, Czech Republic

[*]Contact author: sikola@fme.vutbr.cz

## Abstract

Dielectric metasurfaces offer a path to high-efficiency optical components at the sub-wavelength scale. In this work, we utilize the wavelength-dependent birefringence of $TiO_2$ nanopillars to create transmissive color filters with switchable spectral responses. We demonstrate that the same physical nanostructure array can function as either an additive (band-pass) or a subtractive (band-stop) filter solely by rotating an external polarizer. The metasurface was fabricated via a modified damascene process, enabling high-aspect-ratio structures with spatial resolutions down to 250 nm. The filters achieve transmission efficiencies of 70% tailorable across the entire visible spectrum, avoiding the ohmic losses typical of plasmonic alternatives. This approach provides a tunable, high-efficiency platform offering a versatile and compact solution for next-generation high-resolution imaging systems.

## Introduction

Advances in miniaturization of displays and cameras, notably for VR/AR applications, require increasingly finer spatial resolution of color filters within. Traditional dye-based color filters lose efficiency and color fidelity at the submicron scale, making them unsuitable for the highest-density pixel arrays [1]. This area of extreme spatial resolution combined with thin film requirements is highly suitable for metasurface counterparts. To date, many metasurface color filtering approaches were introduced, each carrying certain advantages and disadvantages connected to their principle of operation.

Early metasurface color filters used metallic nanoantennas and nanohole arrays to produce reflective or transmissive colors via localized plasmon resonances and extraordinary transmission. Although these approaches enabled very small pixel sizes and polarization multiplexing, Ohmic losses in metals typically limit efficiency in the visible range [2–5]. To avoid losses in metals, high-index dielectric metasurfaces ($TiO_2$, Si, $Si_3N_4$, a-Si:H) exploiting the so-called Mie resonances have been developed [6–8]; such devices achieve narrow spectral features with substantially higher transmission or reflection efficiency. Examples such as $TiO_2$ nanopillar arrays and Si/$Si_3N_4$ nanodisks have demonstrated transmissive and reflective colors with off-resonance transmission often higher than 80–90% and substantially improved color purity [9–12]. Beyond mere spectral control, metasurfaces excel in manipulating the polarization state of light through anisotropic or birefringent meta-atoms [13,14].

In this paper, we report on transmissive color filters with high spatial resolution that utilize birefringent properties of a dielectric metasurface to achieve a unique dual-mode operation. This dual-mode functionality is particularly advantageous for high-spatial-resolution image sensors. Unlike displays that require narrow, isolated primary colors, camera subpixels benefit from specific spectral overlaps between RGB channels to accurately infer the impinging spectrum through demosaicing and color correction algorithms. The ability to access both additive and subtractive modes provides a versatile framework for engineering these necessary spectral overlaps at the sub-micron scale.

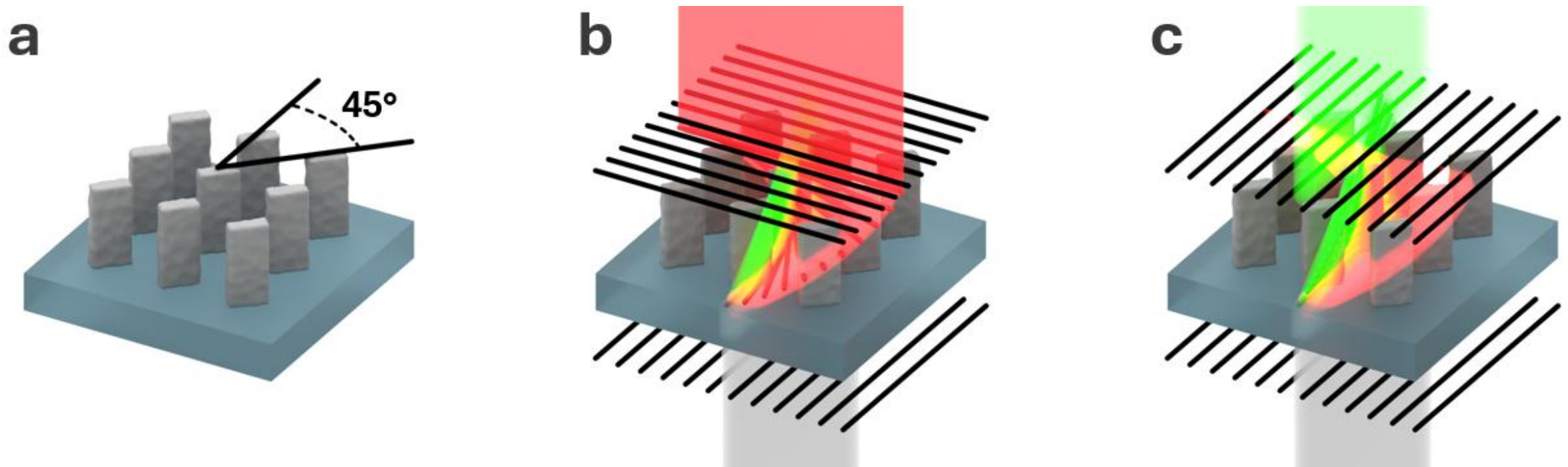

**Fig. 1: Metasurface color filter principle. a)** Nanostructures are rotated 45° within the array lattice. **b)** Metasurface is placed between crossed-oriented polarizers to achieve additive colors **c)** parallel-oriented polarizers or for subtractive colors.

As illustrated in Fig. 1, our device consists of a metasurface sandwiched between two linear polarizers. The metasurface is composed of anisotropic $TiO_2$ nanopillars designed to function as half-wave plates (HWPs) for a specific target wavelength. By orienting these nanostructures at 45° relative to the incident polarization (Fig. 1a), we can rotate the plane of polarization by 90°, allowing light at the target wavelength to pass through a crossed polarizer with a minimal loss (Fig. 1b). Conversely, light polarization at other wavelengths remains largely unrotated and that light is blocked.

The unique feature of this architecture is its inherent switchability: by simply rotating either of the polarizers, the same physical device can switch between an additive (band-pass, Fig. 1b) and a subtractive (band-stop, Fig. 1c) filter. This dual-mode functionality allows for tunable optical responses without the need for complex reconfigurable materials. This approach combines the low-loss benefits of dielectrics with the high spatial resolution of optical metasurfaces, offering a compact and versatile route to high-resolution color filters that complement existing plasmonic and dielectric technologies.

# Methods

## Device principle

The proposed color filter operates on the principle of wavelength-dependent polarization rotation. The device architecture consists of an anisotropic dielectric metasurface sandwiched between two linear polarizers (Fig. 1). The meta-atoms are designed to function as half-wave plates (HWPs) for a specific target wavelength $\lambda_T$. At this target wavelength the nanostructure rotates polarization of incident linearly-polarized light twice the angle between the HWP's fast axis (nanostructure width) and the plane of polarization. By orienting the nanostructures at 45° relative to the polarization plane of the incident $x$-polarized light, the polarization plane is rotated by 90°. This allows the light at $\lambda_T$ to pass through the second (crossed) polarizer with maximal transmission (reduced just by low losses inside the nanostructures). For the light away from $\lambda_T$, however, the nanostructures no longer provide a $\pi$ phase retardance. As a result, the polarization is not fully rotated, and a fraction of the light is blocked by the second polarizer. At wavelengths increasingly detuned from $\lambda_T$, the incomplete polarization rotation causes the second polarizer to block an even larger fraction of the transmitted light. Specifically, the conversion efficiency $T_{xy}$, derived from the Jones matrix formalism [15], for the incident $x$-polarized light to the transmitted $y$-polarized light is given by:

$$T_{xy} = \frac{1}{4}[A_u^2 + A_v^2 - 2A_u A_v \cos(\Delta\phi)], \quad (1)$$

where $A_u$ and $A_v$ are the transmission amplitudes along the two principal axes of the nanostructure, and $\Delta\phi = \phi_u - \phi_v$ represents the phase retardance. In the ideal case of a lossless HWP ($A_u = A_v = 1$ and $\Delta\phi = \pi$), the efficiency $T_{xy}$ reaches 100%, meaning the polarization plane is rotated by exactly $90°$. For wavelengths deviating from $\lambda_T$, the retardance $\Delta\phi$ shifts away from $\pi$, causing the conversion efficiency to drop.

Conversely, in the parallel configuration, the analyzer transmits the $x$-component and blocks the converted $y$-component. This produces a subtractive color filter (band-stop), where a transmission dip appears at $\lambda_T$. As $T_{xx} \approx 1 - T_{xy}$ (assuming no material absorption), the parallel orientation provides a spectral complement to the crossed-polarizer orientation. This inherent symmetry allows a single metasurface design to switch between band-pass filter effect in a crossed-polarizer configuration (additive colors) and a complementary band-stop effect in a parallel configuration (subtractive colors) simply by rotating the external analyzer by 90°.

To realize this principle, we sought anisotropic nanostructures that maximize the $T_{xy}$ for distinct regions of the visible spectrum. These nanostructures will behave like HWPs for the central wavelengths of the given transmission peaks. While our design draws inspiration from Pancharatnam-Berry (PB) metasurfaces - which utilize rotated HWP nanostructures to impose geometric phase and typically operate under circular polarization [16–18] - our primary objective here is to optimize the propagation phase difference (birefringence) to achieve color filtering using standard linear polarizers.

## Fabrication

The color filter nanostructures were fabricated using a method combining electron beam lithography with an atomic layer deposition, called the modified damascene process [16]. First, a fused silica wafer is diced and spin-coated with a resist. We used a positive CSAR resist (AR-P 6200.13) for its good resolution and strong adhesion capabilities. As the thickness of the resist determines the final height of the nanostructures, it is necessary to select a viscous enough resist. The used CSAR 6200.13 results in a 650 nm layer if spin-coated at 1600 RPM. An additional thin layer of chromium (20 nm) was deposited on top to serve as a charge dissipation layer. The nanostructure layout is then transferred into the resist via electron beam lithography. We utilized a 100 kV e-beam system for the exposure, with a dose of 350 μC/cm$^2$. The exposed resist was then chemically stripped of the chromium layer and developed using a standard amyl acetate developer for a duration of one minute. The developed resist mask was then transferred into an atomic layer deposition chamber, where it was conformally coated with a 140 nm $TiO_2$ layer (fully filling the hollow structures fabricated in the resist). Note that ALD film thickness of only half the smallest nanostructure dimension (width) is needed, as the holes in the resist are also filled from the sides. The ALD recipe was optimized for 90 °C not to melt the e-beam resist mask. 2600 cycles of alternating tetrakis(dimethylamino)titanium and water precursors resulted in the 140 nm layer in about 13 hours. The ALD deposition, however, not only fills the resist mask holes but also creates a capping layer over the whole sample surface that needs to be removed. To remove the top layer, we etched the sample surface with a broad argon ion beam in a Scia Systems Coat 200. This system is equipped with a secondary ion mass spectrometer (SIMS) end point detection, which allows us to stop the etching precisely when the top layer is just removed, signified by a titanium peak drop in the SIMS spectra. In the final step, the sample is put into a Diener resist stripper, where the resist mask is removed in oxygen plasma within two hours. This leaves the high-aspect ratio free-standing $TiO_2$ nanostructures with a height of approximately 600 nm on the substrate.

# Results and Discussion

## Optimization of HWP Dimensions

We performed a parametric sweep of the nanostructure dimensions using photonics simulation software, modeling the nanostructures as $TiO_2$ nanopillars on a fused silica substrate with a fixed height of 600 nm. Titanium dioxide was chosen for the platform due to its high refractive index and negligible extinction coefficient in the visible range. To maintain a consistent optical environment and prevent coupling variations between different filter designs, the array pitch (period) was adjusted according to the pillar dimensions to keep the filling factor balanced across all metasurfaces. To identify the most efficient half-wave plate geometries, we mapped the polarization-conversion efficiency $T_{xy}$ across a broad range of lateral nanostructure dimensions $w$ and $l$ at target wavelengths of 450, 550, and 650 nm. Unlike evaluating the phase retardance alone, mapping $T_{xy}$ provides a comprehensive performance metric that accounts for both the phase difference $\Delta\phi$ and the transmission amplitudes $A_u, A_v$ along the principal axes.

As shown in the heatmaps (Fig. 2), the conversion efficiency displays distinct regions of high performance. The "peaks" in these maps correspond to the optimal HWP conditions where the combination of birefringence and high transmission maximizes light rotation. From these maps, we extracted the specific dimensions for our primary color filters (summarized in Table 1), selecting the $(w, l)$ pairs that yielded the absolute highest $T_{xy}$ for each target wavelength. This parametric approach ensures that our selected "best-performing" dimensions are robust against small fabrication tolerances, as they reside within the high-efficiency islands of the design space.

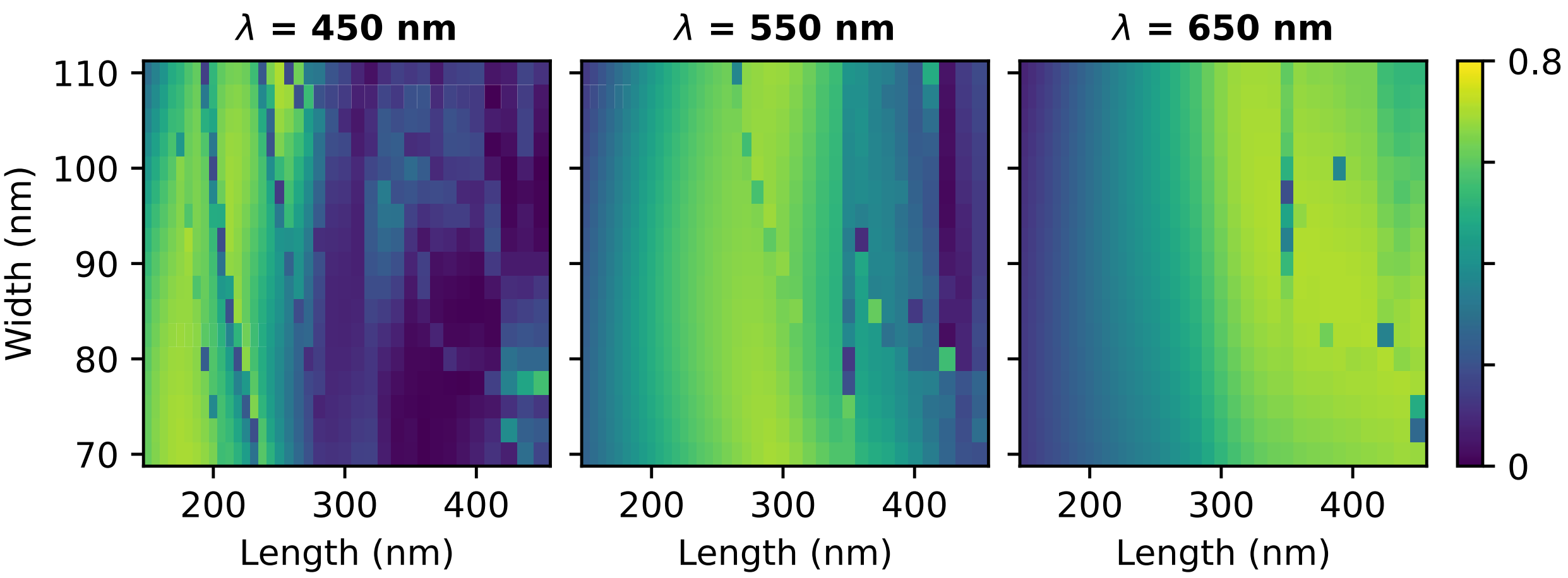

**Fig. 2: Conversion efficiency plots** for nanostructures of varying lateral dimensions at three target wavelengths (450, 550, 650 nm).

## Spectral Response

Using the optimal dimensions identified from the heatmaps, we simulated the broadband transmission spectra to evaluate the bandwidth and efficiency of the filters. Fig. 3 illustrates the $T_{xy}$ and $T_{xx}$ response across the visible range for three representative nanostructure geometries: 90x150 nm (blue light), 80x280 nm (green light), and 80x420 nm (red light).

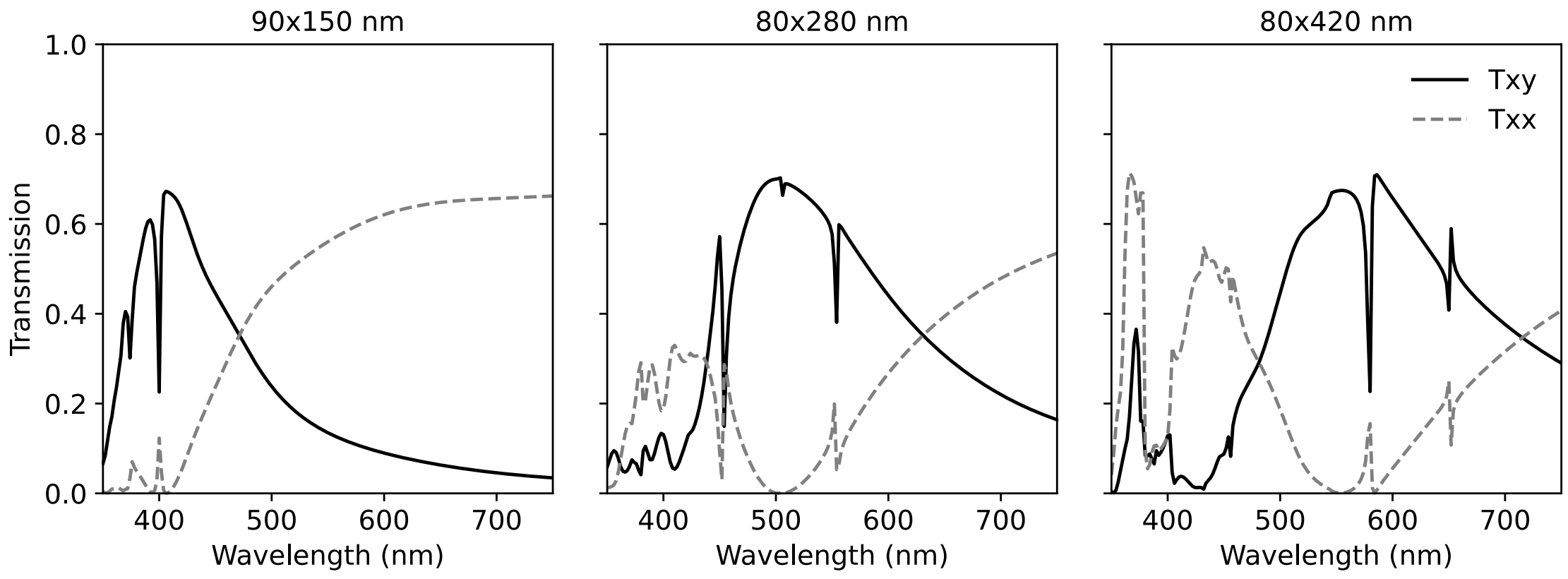

**Fig. 3: Simulated transmission spectra for RGB metasurface filters.** $T_{xy}$ (solid line) and $T_{xx}$ (dashed line) transmission is plotted for three nanopillar geometries: (left) 90x150 nm targeting blue, (center) 80x280 nm targeting green, and (right) 80x420 nm targeting red.

The results demonstrate a clear dual-mode functionality represented by the crossed (solid black line) and parallel (dashed grey line) polarizers configurations. In the crossed-polarizers configuration, the metasurface operates as an additive color filter. For all three designs, well-defined transmission peaks are observed near the target wavelengths, with peak efficiencies reaching approximately 70%. The blue filter exhibits the narrowest bandwidth, while the red filter shows a broader response, consistent with the material dispersion of $TiO_2$ and the wavelength dependence of the effective birefringence, since a fixed spectral shift represents a smaller relative detuning at longer wavelengths. In the context of high-resolution camera sensors, this broadness and the resulting overlap between the RGB transmission curves are intentional and beneficial. This spectral overlap ensures that no 'blind spots' and color ambiguity exist in the visible range, allowing the sensor to mathematically reconstruct the full color gamut of the scene. The side-lobes beyond the main peak in the spectra are minimal, which is crucial for preventing color cross-talk. When the polarizers are set in a parallel configuration, the metasurface instead exhibits a subtractive color response. In these spectra, the transmission reaches a local minimum at the exact resonance wavelength where the crossed-polarizer configuration peaks. Away from these resonance points, the transmission remains high, allowing the majority of the visible spectrum to pass through. The peak efficiency of ~70% for additive colors represents a significant improvement over earlier plasmonic transmissive filters, such as aluminum nanohole arrays, which typically struggle to exceed 15% efficiency due to inherent Ohmic losses [3]. While hybrid metal-on-dielectric systems have recently bridged this gap [19], our all-dielectric $TiO_2$ platform eliminates metallic absorption entirely. Furthermore, the ability to switch to a subtractive mode (band-stop) provides an additional layer of spectral information that can be used to refine the inference of the impinging light's chromaticity.

## Experimental Results

To validate the numerical predictions, arrays of $TiO_2$ nanopillars were fabricated using a modified damascene process combining electron-beam lithography and atomic layer deposition. The fabricated structures span widths from 70 to 110 nm and lengths from 150 to 450 nm, covering the optimized design space. Optical characterization was performed using a custom-built optical setup illustrated in Fig. 4a, consisting of a collimated white-light source, with the metasurface placed between two linear polarizers. The second polarizer (analyzer) was mounted on a motorized rotation stage, allowing rapid switching between crossed and parallel configurations without disturbing the sample alignment. The transmitted light was collected by a spectrophotometer for spectral analysis, and a positioning camera was used to precisely target individual nanostructure arrays.

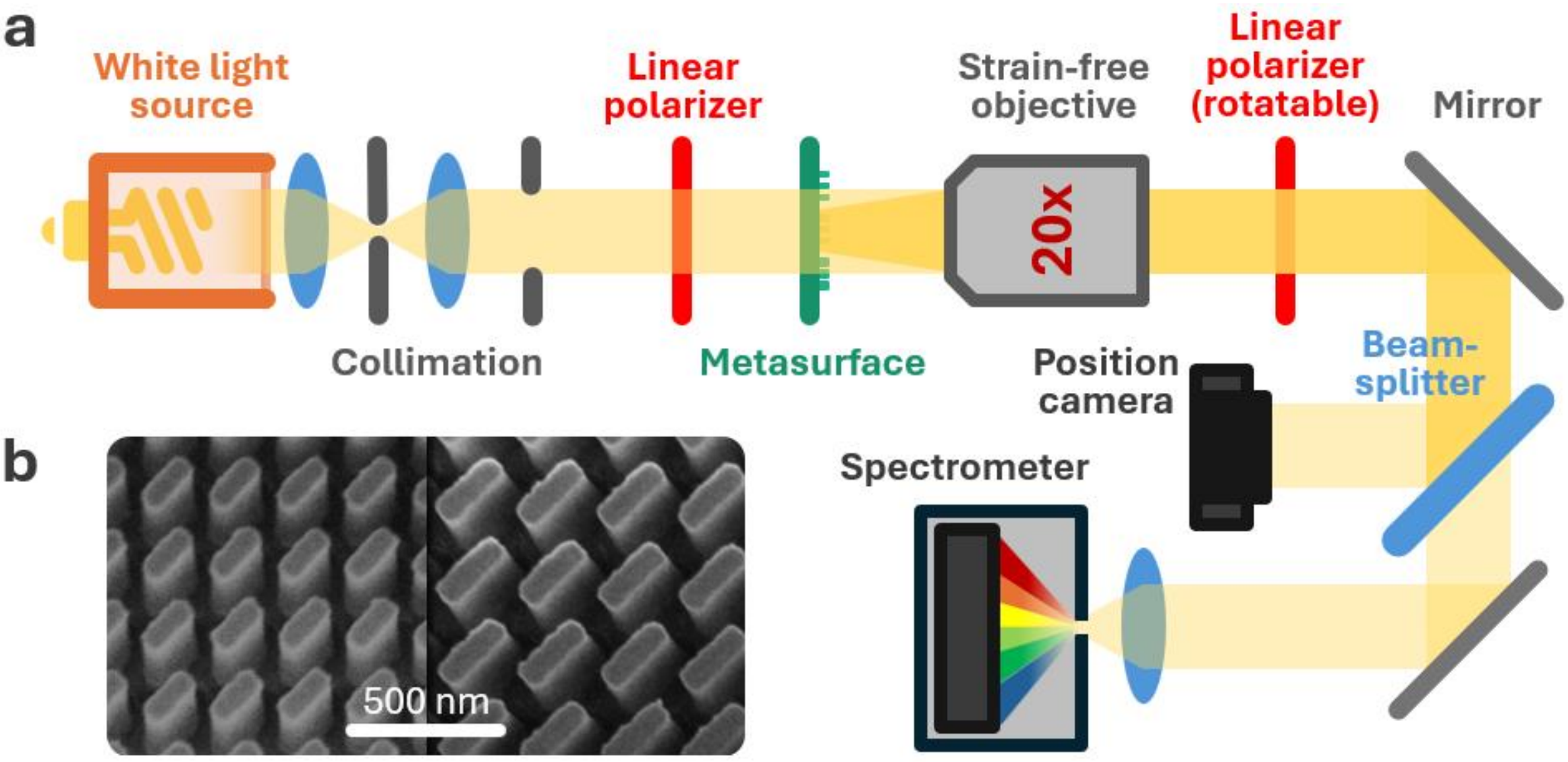

**Fig. 4: a) Optical setup for color filter spectra analysis. b)** Micrograph of fabricated nanostructure arrays.

The experimental performance was evaluated by assigning perceived colors to both simulated and measured spectra by integration over the CIE tristimulus sensitivity functions [20]. The resulting experimental color palettes are compared with the simulated colors in Fig. 5 across a wide range of the fabricated nanostructure sizes. The resulting color palettes show a reasonable agreement between theory and experiment for both additive and subtractive colors. The blue filters are especially notable, achieving high color saturation at a pixel pitch of 250 nm, a regime where conventional dye-based filters typically suffer from severe efficiency loss and color cross-talk. Slight color desaturation observed in some experimental arrays is due to sample-scale fabrication defects that cause broadband depolarized light to leak through the second polarizer. However, the overall consistency between the simulated $T_{xy}$ and $T_{xx}$ peaks and the experimental color output confirms that the device principle is robust. More saturated colors in the longer wavelengths would likely require even taller $TiO_2$ nanostructures. The transition to parallel polarizers (subtractive colors) yielded the expected complementary spectra, further demonstrating the versatile dual-mode operation of our birefringent metasurface.

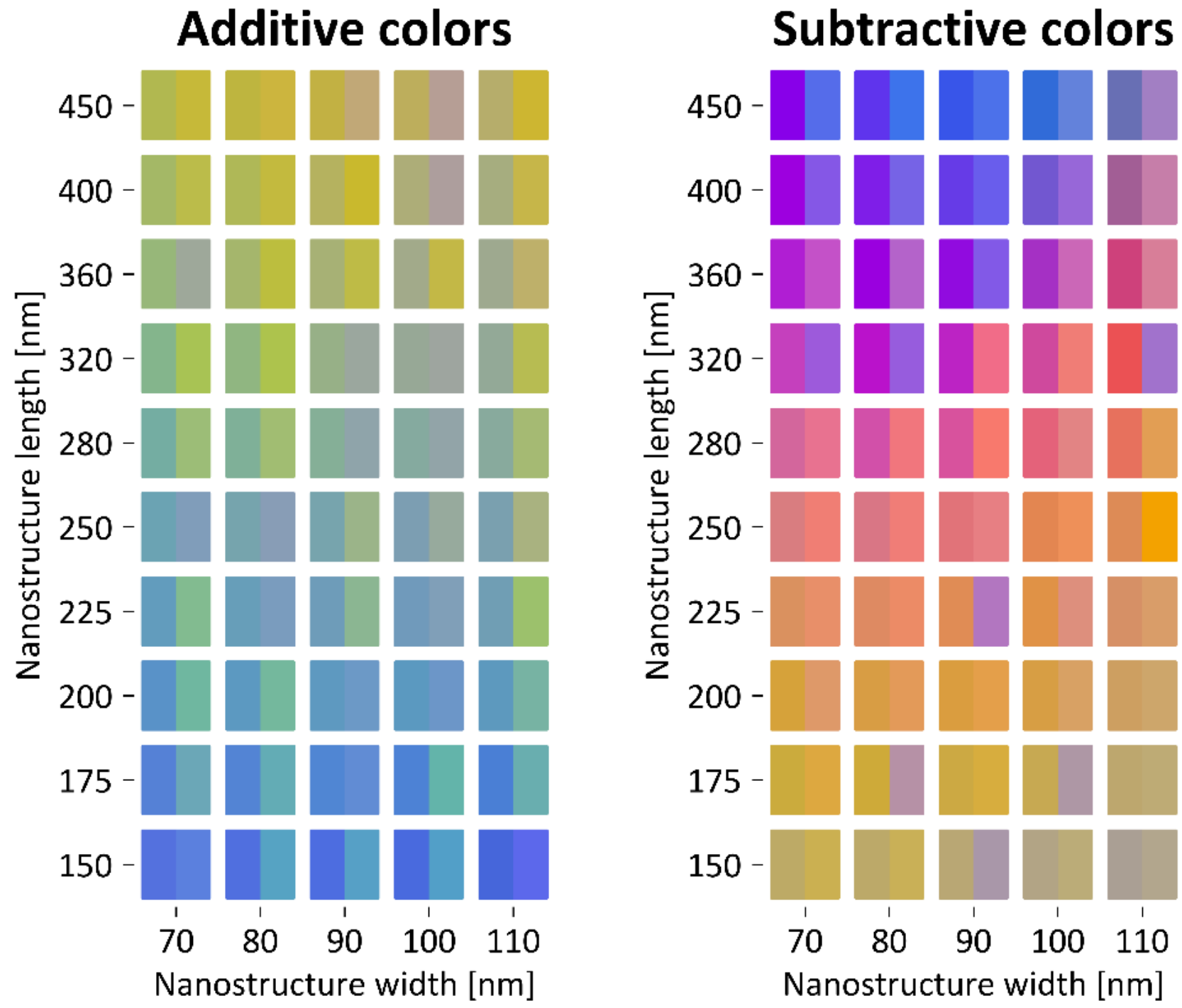

**Fig. 5: Comparison of simulated vs. experimental structural colors.** The additive and subtractive color palettes are shown as a function of nanostructure geometry. Each tile displays a side-by-side comparison of simulated (left half) and experimental (right half) results, both derived from CIE tristimulus integration of the corresponding spectra.

Finally, although the present work focuses on the visible spectral range, the underlying operating principle is not wavelength-specific. The polarization-based filtering relies on achieving a half-wave retardance within an anisotropic nanostructure and analyzing the resulting polarization state using external polarizers. In principle, this approach can be extended to any spectral region for which suitable birefringent nanostructures and polarizing elements can be realized, including the ultraviolet or infrared. The operating wavelength is therefore determined primarily by material dispersion, achievable aspect ratios, and the availability of polarizers, rather than by any fundamental limitation of the device concept.

## Conclusion

We have demonstrated a transmissive color filter architecture based on the birefringent response of anisotropic, high-aspect-ratio $TiO_2$ nanopillars. The peak additive-mode efficiency of our device is approximately 70%, while avoiding metallic absorption. A key feature of the approach is its polarization-controlled dual-mode operation: the same metasurface supports both additive (band-pass) and subtractive (band-stop) filtering, selectable by simple rotation of an external polarizer without modifying the nanostructure geometry. The high fill factor enabled by the nanopillar geometry allows tight spatial resolution down to 250 nm array pitch, and the engineered spectral overlap between color channels is well suited for color inference in high-density imaging systems. These results establish birefringent dielectric metasurfaces as a versatile platform for compact, high-resolution transmissive color filtering.

## Acknowledgement

This work was supported by the Czech Academy of Sciences (Project No. L100652501 and institutional support RVO:68081731), the Czech Science Foundation (Grant No. 23-07617S), and the Technology Agency of the Czech Republic (Project No. TN02000020). The authors also acknowledge the support of Project No. CZ.02.01.01/00/22_008/0004594 (TERAFIT), and CzechNanoLab (Project No. LM2023051) funded by MEYS CR is gratefully acknowledged for the financial support of the measurements/sample fabrication at CEITEC Nano Research Infrastructure.

## Data availability

The data supporting the findings of this study are openly available in the Zenodo repository at https://doi.org/10.5281/zenodo.18481545.

## References

1. R. Rajasekharan, E. Balaur, A. Minovich, S. Collins, T. D. James, A. Djalalian-Assl, K. Ganesan, S. Tomljenovic-Hanic, S. Kandasamy, E. Skafidas, D. N. Neshev, P. Mulvaney, A. Roberts, and S. Prawer, "Filling schemes at submicron scale: Development of submicron sized plasmonic colour filters," Scientific Reports 2014 4:1 4(1), 1–10 (2014).

2. K. Kumar, H. Duan, R. S. Hegde, S. C. W. Koh, J. N. Wei, and J. K. W. Yang, "Printing colour at the optical diffraction limit," Nature Nanotechnology 2012 7:9 7(9), 557–561 (2012).

3. Z. Li, A. W. Clark, and J. M. Cooper, "Dual color plasmonic pixels create a polarization controlled nano color palette," ACS Nano 10(1), 492–498 (2016).

4. S. J. Tan, L. Zhang, D. Zhu, X. M. Goh, Y. M. Wang, K. Kumar, C. W. Qiu, and J. K. W. Yang, "Plasmonic color palettes for photorealistic printing with aluminum nanostructures," Nano Lett. 14(7), 4023–4029 (2014).

5. A. S. Roberts, A. Pors, O. Albrektsen, and S. I. Bozhevolnyi, "Subwavelength plasmonic color printing protected for ambient use," Nano Lett. 14(2), 783–787 (2014).

6. T. Wood, M. Naffouti, J. Berthelot, T. David, J. B. Claude, L. Métayer, A. Delobbe, L. Favre, A. Ronda, I. Berbezier, N. Bonod, and M. Abbarchi, "All-Dielectric Color Filters Using SiGe-Based Mie Resonator Arrays," ACS Photonics 4(4), 873–883 (2017).

7. K. Baek, Y. Kim, S. Mohd-Noor, and J. K. Hyun, "Mie Resonant Structural Colors," ACS Appl. Mater. Interfaces 12(5), 5300–5318 (2020).

8. J. Proust, F. Bedu, B. Gallas, I. Ozerov, and N. Bonod, "All-Dielectric Colored Metasurfaces with Silicon Mie Resonators," ACS Nano 10(8), 7761–7767 (2016).

9. I. Koirala, D.-Y. Choi, and S.-S. Lee, "Highly transmissive subtractive color filters based on an all-dielectric metasurface incorporating TiO2 nanopillars," Optics Express, Vol. 26, Issue 14, pp. 18320-18330 26(14), 18320–18330 (2018).

10. C. S. Park, V. R. Shrestha, W. Yue, S. Gao, S. S. Lee, E. S. Kim, and D. Y. Choi, "Structural Color Filters Enabled by a Dielectric Metasurface Incorporating Hydrogenated Amorphous Silicon Nanodisks," Sci. Rep. 7(1), 1–9 (2017).

11. C.-S. Park, I. Koirala, S. Gao, V. R. Shrestha, S.-S. Lee, and D.-Y. Choi, "Structural color filters based on an all-dielectric metasurface exploiting silicon-rich silicon nitride nanodisks," Opt. Express 27(2), 667 (2019).

12. A. I. Kuznetsov, A. E. Miroshnichenko, M. L. Brongersma, Y. S. Kivshar, and B. Luk'yanchuk, "Optically resonant dielectric nanostructures," Science. 354(6314), 2472 (2016).

13. E. Arbabi, S. M. Kamali, A. Arbabi, and A. Faraon, "Vectorial Holograms with a Dielectric Metasurface: Ultimate Polarization Pattern Generation," ACS Photonics 6(11), 2712–2718 (2019).

14. Y. Yang, W. Wang, P. Moitra, I. I. Kravchenko, D. P. Briggs, and J. Valentine, "Dielectric meta-reflectarray for broadband linear polarization conversion and optical vortex generation," Nano Lett. 14(3), 1394–1399 (2014).

15. R. C. Jones and H. Hurwitz, "A New Calculus for the Treatment of Optical SystemsII. Proof of Three General Equivalence Theorems," JOSA, Vol. 31, Issue 7, pp. 493-499 31(7), 493–499 (1941).

16. R. C. Devlin, M. Khorasaninejad, W. T. Chen, J. Oh, and F. Capasso, "Broadband high-efficiency dielectric metasurfaces for the visible spectrum," Proc. Natl. Acad. Sci. U. S. A. 113(38), 10473–10478 (2016).

17. S. Choudhury, U. Guler, A. Shaltout, V. M. Shalaev, A. V Kildishev, A. Boltasseva, S. Choudhury, U. Guler, V. M. Shalaev, A. V Kildishev, A. Boltasseva, and A. Shaltout, "Pancharatnam–Berry Phase Manipulating Metasurface for Visible Color Hologram Based on Low Loss Silver Thin Film," Adv. Opt. Mater. 5(10), 1700196 (2017).

18. Q. Deng, J. Yang, X. Lan, W. Zhang, H. Cui, Z. Xie, L. Li, and Y. Huang, "Investigations of generalized Pancharatnam-Berry phase in all-dielectric metasurfaces," Results Phys. 51, 106730 (2023).

19. C. Lee, S. Lee, J. Seong, D. Y. Park, D. Y. Park, J. Rho, J. Rho, J. Rho, J. Rho, and J. Rho, "Inverse-designed metasurfaces for highly saturated transmissive colors," JOSA B, Vol. 41, Issue 1, pp. 151-158 41(1), 151–158 (2024).

20. J. Morovic and J. Lammens, "Colorimetry: Understanding the CIE system," 159–206 (2007).